\documentclass[aps,prl,showpacs,twocolumn,groupedaddress]{revtex4}
\usepackage[latin1]{inputenc}
\usepackage{graphicx}
\usepackage{color}

\begin{document}


\title{Dynamics of cubic-tetragonal phase transition in KNbO$_3$ perovskite}



\author{S.N. Gvasaliya}
\altaffiliation{On leave from Ioffe Physical Technical Institute, 
26 Politekhnicheskaya, 194021, St. Petersburg, Russia}
\affiliation{Laboratory for Neutron Scattering ETHZ \& Paul-Scherrer
Institut CH-5232 Villigen PSI Switzerland}

\author{B. Roessli}
\affiliation{Laboratory for Neutron Scattering ETHZ \& Paul-Scherrer
Institut 
CH-5232 Villigen PSI, Switzerland}

\author{R.A. Cowley}
\affiliation{Clarendon Laboratory, Oxford University, Parks Road, Oxford OX1 3PU, UK}

\author{S.G. Lushnikov}
\affiliation{Ioffe Physical Technical Institute, 26 Politekhnicheskaya,
194021, St. Petersburg, Russia}

\author{A. Choubey}
\affiliation{Nonlinear Optics Laboratory, ETH Honggerberg, CH-8093 Z\"urich, Switzerland}

\author{P. Gunter}
\affiliation{Nonlinear Optics Laboratory, ETH Honggerberg, CH-8093 Z\"urich, Switzerland}

\date{\today}

\begin{abstract}
The low-energy part of the vibration spectrum in KNbO$_3$ was studied 
by cold neutron inelastic scattering in the cubic phase. In addition to acoustic 
phonons, we observe strong diffuse scattering, which consists of two components. 
The first one is quasi-static and has a temperature-independent intensity. The 
second component appears as quasi-elastic scattering in the neutron spectrum 
indicating a dynamic origin. From analysis of the inelastic data we conclude 
that the quasi-elastic component and the acoustic phonon are mutually coupled.  
The susceptibility associated with the quasi-elastic component grows as the 
temperature approaches T$_C$.
\end{abstract}


\maketitle

\noindent 

ABO$_3$ perovskites form a class of important materials, in part because 
of potential technical applications but also as fundamental interest 
in the physics of phase transitions [1,~2]. At sufficiently high 
temperatures many of these perovskites have O$_h^1$ cubic symmetry and    
structural phase transitions can take place as the temperature is lowered. 
Well-known examples are {\it e.g.} the cubic-tetragonal phase transition 
in SrTiO$_3$ ($T_C\approx105$~K) or in BaTiO$_3$ ($T_C\approx425$~K) 
(for a review see Ref.~[3]).  There are, however, ABO$_3$ perovskites 
which were less studied. An example is the first-order 
cubic-tetragonal phase transition in KNbO$_3$ which occurs 
at T$_{C}$~$\approx683$~K when cooling the crystal from above the 
transition temperature~[4].              

The mechanism of the cubic-tetragonal (CT) phase transition in KNbO$_3$ is 
still controversial.  Whereas well-defined soft phonon modes with 
frequency varying with temperature have been detected in many  
materials close to T$_C$ [1~-~3], only an over-damped excitation 
has been observed in cubic KNbO$_3$ with neutron scattering and it was 
suggested that the nature of the C-T phase transition in that compound 
is similar to the displacive C-T transition in BaTiO$_3$~[5,~6]. On the other hand, two 
coexisting and essentially uncoupled modes are inferred from analysis of 
optical data in the cubic phase of KNbO$_3$: a relaxation mode and a 
soft phonon, with the relaxation process driving the C-T phase transition~[7]. 

We re-investigated the low energy part 
of the vibration spectrum in KNbO$_3$ under improved resolution conditions
first to try to elucidate the mechanism of the phase transition in 
this crystal and second to check whether the diffuse scattering found in Ref.[8] 
is of static or dynamic origin. The inelastic cold-neutron scattering 
measurements reported here were performed with the three-axis spectrometer 
TASP, located at the neutron spallation source SINQ (Paul Scherrer Institute, 
Switzerland). A large single crystal of KNbO$_3$ 
($\sim$ 20~cm$^3$, mosaic $\sim 80'$) was mounted into an ILL-type furnace. 
To decrease the level of incoherent background the sample  
holder was made from pure niobium. The crystal was aligned in the (h~k~0) 
scattering plane. The measurements were performed in the temperature range 
$727$~K - $1030$~K. The (002) reflection of pyrolytic graphite (PG) was used 
to monochromate and analyze the incident and scattered neutron beams, 
respectively. The spectrometer was operated in the constant final-energy 
mode with $k_{f}=1.97 {\rm \AA}^{-1}$. A PG filter was used to remove 
higher-order wavelengths. The horizontal collimation was 
$10'/{\rm \AA}-80'-80'-80'$. With that configuration the energy resolution 
at zero energy transfer is $\sim$~0.4 meV. 
By monitoring the position and intensity of the (1,~1,~0) Bragg
peak, the temperature of the cubic-tetragonal phase transition upon cooling 
was found at $T_C=684\pm2$~K, in close agreement with published data [4,~6].   
The temperature of the sample was controlled by two thermocouples. The temperature 
gradient through the sample did not exceed 15~K. 

%
%
\begin{figure}[h]
  \includegraphics[width=0.5\textwidth, angle=0]{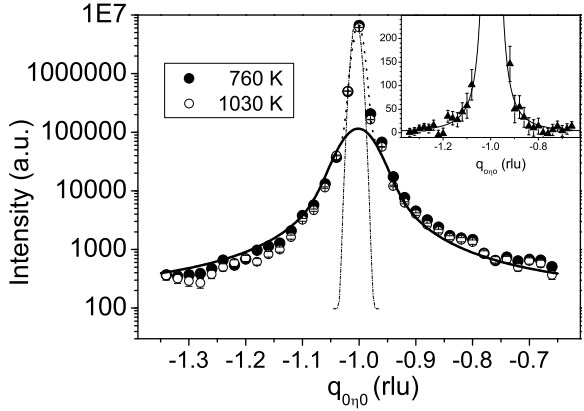}
  \caption{Profiles of elastic scans in the (0$\eta$0) 
           direction at T~=~760~K and T~=~1030~K . Raw data are shown 
           by circles. The dotted line is the result of fits as described 
           in the text. The bold line shows the Lorentzian profile, 
           the dashed-dotted line stands for the intense and narrow 
           Bragg peak. The intensity is given in a logarithmic scale. 
           The insert shows the difference of the elastic scans 
           $\rm I(T=760K)$-$\rm I(T=1030K)$ fitted with Eq.~\ref{lor}. 
           1 rlu corresponds to 1.57 $\rm\AA^{-1}$. 
           } 
\label{fig1}
\end{figure}

\begin{figure}[h]
  \includegraphics[width=0.5\textwidth, angle=0]{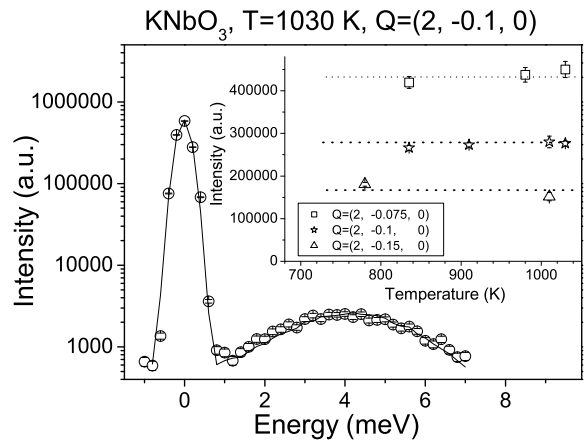}
  \caption{Neutron scattering spectrum from KNbO$_3$ at 1030 K. Raw data 
           are shown by open circles. The solid line is the result of fit as 
           described in the text. The intensity is given in a 
           logarithmic scale. The apparent width of the phonon peak is 
           due to the resolution effects. The insert shows the temperature 
           and the q dependences of the intensity of the central peak. 
           1 rlu=1.567 $\AA^{-1}$.}
\label{fig2}
\end{figure}

Before analyzing the inelastic neutron spectra quantitatively, it is 
convenient to address the q-dependence of the elastic neutron response. 
Figure~\ref{fig1} shows representative elastic scans along the 
(1,1$\pm$q,0) direction at T=760~K and 1030~K, respectively. This 
intense and broad scattering is similar to the diffuse scattering 
observed in KNbO$_3$ by Guinier et al.~[8] using X-rays and reflects the  
presence of atomic disorder in the perovskite cell. In KNbO$_3$ atomic 
disorder yields diffuse scattering along the [100] direction both 
in the X-ray and neutron diffraction patterns. Here we approximate 
the line-shape of the neutron diffuse scattering intensity by a 
Lorentzian profile: 
\begin{equation}
\label{lor}
A(q)= {1\over{\pi}}\frac{I_0}{(q-q_0)^2+\kappa^2}
\end{equation}
\noindent where $q_0$ is the position in reciprocal space; $\kappa$ 
the inverse of the correlation length $\xi$ and $I_0$ yields the integrated 
intensity. From a fit to the elastic data at T~=~1030~K we obtain 
$\xi=64\pm6\rm~\AA$. It turns out that the shape and intensity of the diffuse 
scattering measured in the (2,0,0) Brillouin zone (BZ) does not depend on temperature 
(see insert of Fig.~\ref{fig2}).  This is in agreement with the results of Ref.~[8] where 
the intensity of the diffuse scattering is found to be temperature independent in the 
cubic phase and to decrease abruptly by $\sim30\%$ immediately below T$_C$. 
On the other hand, it turns out that the intensity of the diffuse scattering measured 
along (1,1$\pm q$,0) slowly decreases when increasing the temperature from $T_C$. 
This suggests that in that BZ and for temperatures relatively close to 
$T_C$, the diffuse scattering consists of two Lorentzian components 
(see inset of Fig.~1). 

We turn now to the analysis of the inelastic neutron scattering spectra. 
Figure~\ref{fig2} shows an example of a constant-q scan taken 
at ${\bf Q}=$(2,~-0.1,~0) and T~=~1030~K. The spectrum 
contains an inelastic peak at $\hbar\omega$=4~meV from the transverse 
acoustic (TA) phonon and a narrow peak centered around zero energy transfer. 
To analyze the data quantitatively, we hence 
modeled the neutron scattering intensity $I(\mathbf{Q},\omega)$ in the 
following way: 
\begin{equation}
\label{f1}
I(\mathbf{Q},\omega)=S(\mathbf{Q},\omega){\otimes}\,R(\mathbf{Q},\omega)+B
\end{equation}
The symbol $\otimes$ stands for the 4-D convolution with the spectrometer 
resolution function $R(\mathbf{Q},\omega)$~[9]; $B$ denotes the background level; 
$S(\mathbf{Q},\omega)$ is the neutron scattering function which is related to the the 
imaginary part of the dynamical susceptibility $\chi''(\mathbf{Q},\omega)$  through          
\begin{equation}
\label{sqw2}
S(\mathbf{Q},\omega)={[n(\omega)+1]\over{\pi}}\chi''(\mathbf{Q},\omega) 
\end{equation}
with the temperature factor $[n(\omega)+1]=[1-\exp(-\omega/T)]^{-1}$. 
We approximate the central peak by a $\delta$-function in energy 
\begin{equation}
 \label{CP}
 S_{CP}=A({q})\delta(\omega)
 \end{equation}
with $A({q})$ given by Eq.~\ref{lor}.  
The line-shape of the acoustic phonon is given by the usual
damped-harmonic oscillator (DHO) 
 \begin{equation}
 \label{dho}
  \chi_{DHO}({\bf q},\omega)=(\Omega^2_q-i\gamma_q\omega-\omega^2)^{-1}.
 \end{equation}
\noindent  In Eq.~\ref{dho},
 $\gamma_q$ is the damping and   
$\Omega_{q}=\sqrt{\omega_{q}^2+\gamma_{q}^2}$ with $\omega{_q}=c\cdot q$~[10] 
is the renormalized frequency of the acoustic phonons. For small values of momentum 
transfers q, a linear dispersion for the acoustic phonon branch is a reasonable 
approximation and the phonon damping approximatively follows a $dq^2$-dependence~[11].
The scattering function used to fit the neutron data then reads  
\begin{equation}
S(\mathbf{Q},\omega)=S_{CP}({\bf Q},\omega) +\frac{[n(\omega)+1]}{\pi}f_1^2\chi_{DHO}''({\bf Q},\omega)
\label{sqwx}
\end{equation}
where $\mathbf{Q}=\mathbf{q}+{\bf \tau}$ is the neutron scattering vector and $\bf \tau$ 
a reciprocal lattice vector; $f_1$ is the structure factor of the acoustic phonon. 
As shown in Fig.\ref{fig2}, Eqs.~1-6 parameterize the 
experimental data in the (2,~0,~0) BZ well. The central peak is resolution-limited 
and temperature-independent. The acoustic phonon branch has a stiffness 
$c=28\pm1.3$~meV$\cdot\rm\AA^2$ and the damping is small at low q, 
$\gamma_q=dq^2$ with $d=55\pm 6$ meV$\rm\AA^2$ (0.05$<q<0.15$ (rlu)). 
In the temperature range 750~K$<T<$1030~K no qualitative change in the dispersion 
of the acoustic phonon was observed for data taken along (2,~q,~0) ($|q|<$0.15
(rlu)). 

\begin{figure}[h]
  \includegraphics[width=0.5\textwidth, angle=0]{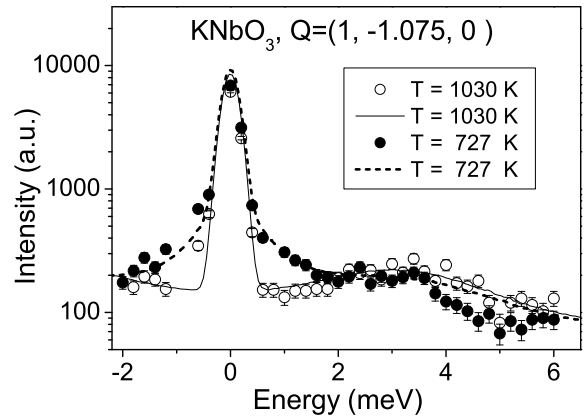}
  \caption{Neutron scattering spectrum from KNbO$_3$ at T=1030 and 727
           K, respectively. The solid and dashed line are the results of fit as 
           described in the text. To emphasize the QE component the 
           intensity is given in a logarithmic scale. Note the pronounced change 
	   in the phonon line-shape at lower temperature.}
\label{fig3}
\end{figure}

On the contrary, for temperatures below T=1030~K an additional component 
is observed in the inelastic spectra for constant-q scans in the (1,1,0) BZ. 
For means of comparison, Fig.~\ref{fig3} shows two representative neutron scattering 
spectra measured in the (1,~1,~0) BZ at T=1030~K and T=727~K, respectively. 
At T~=~1030~K the spectrum consists of two components - a 
central peak (CP) and a phonon response around $\hbar\omega=$4 meV. 
However, as the temperature is lowered to T=727~K, additional quasielastic 
scattering (QE) appears along (1,~1$\pm q$,~0) direction.  The
intensity of this quasi-elastic scattering grows when approaching T$_C$.
%
%
%
%
From the above  discussion we conclude that in the (1,~1,~0) BZ, 
the inelastic neutron spectra consist of three contributions: a central peak, 
quasi-elastic and phonon scattering. To describe the quasi-elastic scattering we 
introduce a Debye-like relaxation function
 \begin{equation}
 \label{quasiel}
 \chi_{q-el}({\bf q},\omega)=\frac{\chi(0,T)}{1+q^2/\kappa^2}\cdot(1-i\omega/\Gamma_q)^{-1}, 
 \end{equation}
$\chi(0,T)$ is the temperature dependent static susceptibility;
$\kappa$ the inverse of the correlation length and $\Gamma_q= \Gamma_0+Dq^2$. 
Taking into account the quasi-elastic scattering modifies the neutron cross-section to 
\begin{eqnarray}
\label{sqw1}
S(\mathbf{Q},\omega)&=&S_{CP}({\bf Q},\omega)  \\ 
&+&\frac{[n(\omega)+1]}{\pi}[f_1^2\chi_{DHO}''({\bf Q},\omega)+f_2^2\chi_{q-el.}''({\bf Q},\omega)], \nonumber
\end{eqnarray}
However, the scattering function given in Eq.~\ref{sqw1} fails in reproducing the
experimental data in the (1,~1,~0) BZ for T$<$1030~K. For example Fig.~\ref{fig3} 
shows an inelastic spectrum at T=727~K and ${\bf Q}$=(1,~1.075,~0) where a 
qualitative change in the phonon line-shape accompanied by a shift in the position 
of the phonon peak is observed. These two effects suggest that coupling between the 
quasi-elastic component and the acoustic phonons becomes important as the temperature 
approaches T$_C$. 
%
%
%
\begin{figure}[h]
\includegraphics[scale=0.75, angle=0]{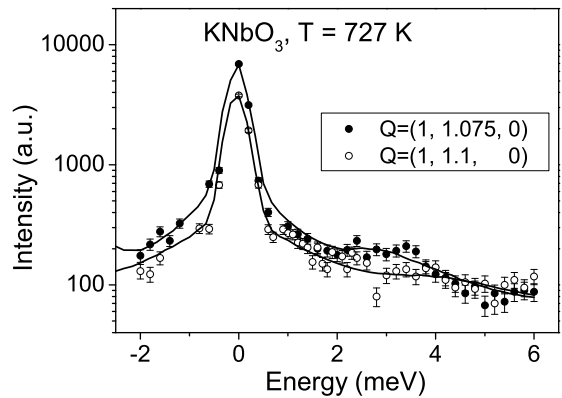}
\caption{Observed and fitted inelastic neutron intensities taken 
         in the (1,~1,~0) BZ at T=727~K. Fitted curves were obtained 
         with Eq.~\ref{sqw3}. Intensity is given in a logarithmic scale.}
\label{fig4}
\end{figure}

\begin{figure}[h]
\includegraphics[scale=0.75, angle=0]{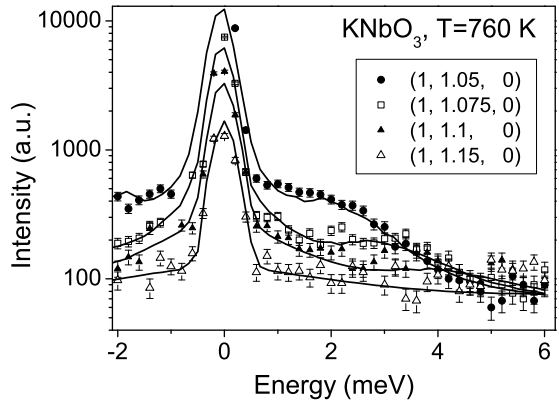}
\caption{Observed and fitted inelastic neutron intensities taken 
         in the (1,~1,~0) BZ at T=760~K. Fitted curves were obtained 
         with Eq.~\ref{sqw3}. Intensity is given in a logarithmic scale.}
\label{fig5}
\end{figure}

The dynamical susceptibility for two coupled excitations was considered 
in details in Refs.~[3,~12,~13] and is given by 
%
\begin{equation}
\label{ccm}
\chi_{CM}({\bf Q},\omega)=\frac{f_1^2\chi_1+f_2^2\chi_2+2\lambda f_1f_2\chi_1\chi_2}
{1-\lambda^2\chi_1\chi_2}
\end{equation}
where $\chi_i\equiv\chi_i({\bf Q},\omega)$, $i=1,2$ are the dynamical 
susceptibilities of the uncoupled phonon and QE component, respectively.
In the following we take $f_i$ as real constants since in KNbO$_3$ all the atoms 
are situated on centers of symmetry. The interaction term is
$\lambda\equiv\lambda(q,\omega)=(g_r+i\omega{g_i})\,q^2$. 
Finally, the scattering function reads: 
%
\begin{equation} 
\label{sqw3} 
S(\mathbf{Q},\omega)=S_{CP}({\bf Q},\omega)+
\frac{[n(\omega)+1]}{\pi}\chi''_{CM}.
\end{equation}  
In order to obtain a good agreement between Eq.~\ref{sqw3} and the neutron 
spectra, it was necessary to fit the complete set of data (0$<$q$<$0.2) taken 
at a given temperature simultaneously. Figures~\ref{fig4} and ~\ref{fig5} show 
the results of such calculations for T=727~K and T=760~K. We obtain 
$g_r=20\pm 3$~meV$^2\cdot\rm\AA^2$ and $g_i=95 \pm 6$ meV$\cdot\rm\AA^2$.
Introduction of a coupling between the QE and acoustic modes has two consequences. 
First, Eq.~\ref{ccm} yields a better description of the line-shape of the inelastic 
neutron spectra. Second, $\chi_{CM}''({\bf Q},\omega)$ is enhanced at low energy 
transfers. Further, we obtained $\Gamma_0=0.19\pm0.05$~meV and 
$D= 44 \pm 4$~meV$\cdot\rm\AA^2$ for the damping of QE component. 
%
%
%
%
%
%
%
%
\begin{figure}[h]
\includegraphics[scale=0.75, angle=0]{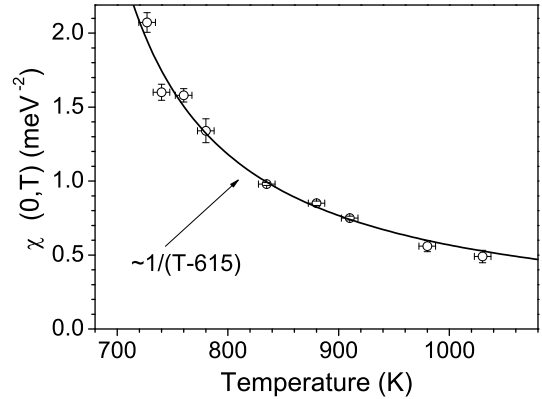}
\caption{Temperature dependence of the susceptibility of the QE 
         component. The solid line is a fit to the data as explained 
         in the text.
	}
\label{fig6}
\end{figure}
%
%
As discussed above, both the CP and the line-shape of the acoustic phonons are 
temperature independent in the (2,~0,~0) BZ. Hence, to fit the data measured in
the (1,~1,~0) BZ as a function of temperature, we fixed the parameters of 
the CP and the acoustic phonons. 
The only parameter left to describe the temperature dependence of the neutron 
spectra is the susceptibility of QE scattering $\chi(0,T)$. 
As shown in Fig.~\ref{fig5} the intensity of the quasi-elastic component 
is a maximum close to T$_C$  and decreases continuously with increasing 
temperature. The temperature dependence of $\chi(0,T)$ follows approximately 
the Curie-Weiss law $\propto 1/(T-T_0)$ with $T_0=615\pm23$~K in good agreement 
with the value deduced from  dielectric measurements $T_0=633\pm5.9$~K 
(Ref.~[14]), and $T_0=615$~K (Ref.~[15]). 
This suggests that the cubic-tetragonal transition in KNbO$_3$ is driven 
by the quasi-elastic relaxational excitation. The intensity of the quasi-elastic component 
is strong in the (1,1,0) zone and has a small intensity in the (2,0,0) Brillouin
zone, which indicates that the relaxation mode is due to correlated atomic motion of 
optical character. However, at all q and temperatures we did not 
observe that the relaxation mode evolves into an under-damped optic phonon
branch. 
Thus, we conclude that QE scattering in KNbO$_3$ is not due to 
an usual overdamped soft phonon but is related to disorder in the lattice. 

To summarize, we measured the low-energy part of the vibration spectrum of 
KNbO$_3$ in the cubic phase with inelastic neutron scattering. We find a   
coexistence of a static and a quasi-elastic component. The static component 
appears to correspond with static disorder in the cubic cell and is temperature 
independent in agreement with X-rays results~[5]. The quasi-elastic component 
is coupled with the acoustic phonon branch and its intensity follows 
the Curie-Weiss law well.  

\begin{acknowledgments}

This work was performed at the spallation neutron source SINQ, 
Paul Scherrer Institut, Villigen (Switzerland) and was partially 
supported by RFBR grant 02-02-17678. P. G\"unter and A. Choubey acknowledge
partial support by the Swiss National Science Foundation. 

\end{acknowledgments}

%

\end{document}